\documentclass{elsart}

\def\beq{\begin{equation}}
\def\enq{\end{equation}}
\def\bearr{\begin{eqnarray}}
\def\enarr{\end{eqnarray}}
\def\C{{\mathcal C}}
\def\I{{\mathcal I}}
\def\R{{\mathcal R}}

\def\n{\noindent}

\def\ie{{i.e.}}

\def\JSP{{J.\ Stat.\ Phys.}\ }
\def\JPA{{J.\ Phys.\ A}\ }
\def\PRL{{Phys.\ Rev.\ Lett.}\ }
\def\PRA{{Phys.\ Rev.\ A}\ }
\def\PRE{{Phys.\ Rev.\ E}\ }
\def\PLA{{Phys.\ Lett.\ A}\ }
\def\PFA{{Phys.\ Fluids A}\ }
\def\PhysicaD{{Physica D}\ }
\usepackage{psfig}

\begin{document}

\begin{frontmatter}

\title{Lacunarity of Random Fractals}
\author[newton,franck]{Francisco J. 
Solis\thanksref{NWU}} and
\author[damtp]{Louis Tao\thanksref{corr}}
\address[newton]{The Isaac Newton Institute of Mathematical Sciences,
20 Clarkson Road, Cambridge CB3 0EH, United Kingdom}
\address[franck]{The James Franck Institute, University of Chicago,
5640 S. Ellis Avenue, Chicago, IL 60637, USA}
\address[damtp]{Department of Applied Mathematics 
and Theoretical Physics,
University of Cambridge, Silver Street,
Cambridge CB3 9EW, United Kingdom}
\thanks[NWU]{Present address: Department of
Materials Science and Engineering,
Northwestern University, Evanston, IL 60208.
E-mail: solis@ren.ms.nwu.edu}
\thanks[corr]{Corresponding author.
E-mail: L.L.Tao@damtp.cam.ac.uk}

\date{\today}

\maketitle
\begin{abstract}
We discuss properties of random fractals by means of a set of numbers
that characterize their universal properties. This set is the 
generalized singularity spectrum that consists of the usual spectrum
of multifractal dimensions and the associated complex analogs. 
Furthermore, non-universal properties are recovered from the study of
a series of functions which are generalizations of the so-called
energy integral.
%singular points, which correspond to the generalized singularity spectrum.

\end{abstract}

%\n{PACS numbers: 47.53.+n, 0.2.50.-r}

\begin{keyword}
Fractals, stochastic processes
\end{keyword}

\end{frontmatter}

A number of the properties of fractals are associated with their
Hausdorff, box-counting, or fractal dimensions. But further
information of a universal character is also encoded in secondary
(singular) dimensions, some of which may be complex.
For example, a recurrent theme in the study of fractals is that of 
asymptotic or logarithmic periodicity \cite{Badii,Smith}.
Most fractal objects and mathematical
constructions thereof are not exactly scale invariant. Rather, they
obey simple recurrence relations that relate an infinite but
discrete set of scales.  Recent physical examples include
the appearance of complex exponents in diffusion-limited
aggregation \cite{Saleur}, crack propagation in two dimensions
\cite{Blumenfeld}, and Boolean delay equations in the modelling of
climate dynamics \cite{Mullhaupt}. 

For instance, instead of the full
scale invariance of a function of local variables,
\beq
	F(x)=\lambda^{D}F(x/\lambda),
\label{eq:scaling}
\enq
we have the logarithmic analog of Bloch's theorem,
\beq
  F(x)=\lambda^{D}F(x/\lambda)\, G(\ln \lambda),
%  F(x)=\lambda^{D}F(x/\lambda)\, \frac{G(\ln x)}{G[\ln (x/\lambda)]},
\label{eq:bloch}
\enq
where $G$ may be a periodic function. 

This is a rather well-known relation but we feel that it has
been under exploited.  The period of $G$ is independent 
of the scaling dimension, $D$, and gives further information
of the properties of the fractal object.

The ways in which the fractal dimensions of the object manifest itself 
are manifold.  Consider, in particular, a set of analytical quantities 
calculated from the real space distribution of a 
fractal object, namely, the set of correlation integrals:
\beq
  C_q(r)
    = \int d\mu(x) \left(\int d\mu(y) \,\,\theta(r - |x - y|)\right)^{q-1},
\label{eq:correlation}
\enq
where $q \ge 2$, $\theta$ is the Heaviside step function, and 
$d\mu$ is the measure of the object in question.  The
scaling properties of these correlation integrals with respect to
the distance $r$ define a countable set of dimensions forming
part of the multifractal dimension spectrum
\cite{GP,Halsey,PV}.

The study of the scaling properties of such correlations is
facilitated by considering the corresponding energy integrals:
\beq
  I_q(\tau) = \int r^{-\tau}\, d C_q(r), \quad \tau < d,
\label{eq:energy}
\enq
where $\tau$ is restricted to be less than the spatial 
dimension, $d$.  Note that $I_q(\tau)$ is related to the
Mellin transform of $C_q(r)$ \cite{Mellin}.

At this point the spatial information of the fractal has been encoded
in these energy integrals, which are typically, but not always,
meromorphic.  For particular cases of
deterministic Cantor sets, it is shown in \cite{Bessis,Fournier} that
the complex structure of these functions (\ref{eq:energy}) reveal
singularities that correspond to the relevant scaling dimensions
of the theory.  One usually one keeps only the most relevant,
\ie, the one with the smallest real part, but the rest of the
spectrum is important in studying finite-size effects.

More precisely, the most relevant singularity is a pole on
the real axis and has a numerical value that is a lower bound
to the Hausdorff dimension.  This was first proved
by Frostman for $q = 2$ \cite{Frostman} (see also
Falconer \cite{FalconerBk}).  % In Frostman's terminology,
% $I_2(\tau)$ is called the $\tau$-energy of the fractal measure.
In some cases, the Hausdorff dimension also corresponds to
the box-counting dimension \cite{FalconerBk,Mainieri}.

Typically, the rest of the singularities are also poles, and
appear as pairs of complex conjugates with real parts not
smaller than the Hausdorff dimension.
The imaginary parts of these poles correspond obviously
to the logarithmic wavelength of the fractal, while the 
residues appear as the amplitudes
of oscillations observed in the asymptotic scaling of various
correlation integrals (\ref{eq:correlation}).  This program of
the analysis of a fractal object has been carried out, albeit in
a somewhat scattered way, for the middle-third Cantor set and
some of its deterministic generalizations 
\cite{Bessis,Fournier,Orlandini}.
This complex singularity spectrum has been called the
{\it multilacunarity} spectrum \cite{Fournier}.  It is the
goal of our work to show that it is well-defined for classes
of random fractals, and we explicitly compute the lacunarity
of a particular example.

It turns out that for some simple but important examples
the Hausdorff dimension is easy to calculate
with the use of a little ingenuity.  Just as simply,
the satellite dimensions can be calculated in the same way
without resorting to the explicit computation of the correlation 
integrals (\ref{eq:correlation}) or the energy integrals 
(\ref{eq:energy}).

For the middle-third Cantor set, and for many other objects with
simple recursive descriptions, %\cite{markov}, we consider
we consider
an equation that relates the relative scales, $\ell_i$, and the
relative (normalized) measures, $p_i$, at successive levels of
approximation,
\beq
\sum_i \frac{p_i^q}{\ell_i^{\tau}} = 1.
\label{eq:partition}
\enq

In the case of the middle-third
Cantor set, $\ell_{i=1,2} = 1/3$ and $p_{i=1,2} = 1/2$, 
and (\ref{eq:partition}) becomes $2^{q-1}/{3^\tau} = 1$.
%\beq
%\frac{2^{q-1}}{3^\tau} = 1.
%\label{eq:middlethird}
%\enq
The unique real solution, $\tau(q) = (q-1) \ln 2 / \ln 3$, 
gives the Hausdorff dimension.  In general, $\tau=\tau(q)$
is not linear in $q-1$ and gives one the desired multifractal
dimension spectrum via the relation $D_q = \tau(q)/(q-1)$.
In this way, the universal properties of the fractal, namely,
its generalized dimensions, are rather easily
computed \cite{Halsey,Feigenbaum}.

However, as noticed in \cite{Fournier}, a study of
the middle-third Cantor set (and some deterministic
generalizations), (\ref{eq:partition}) also has complex roots.
For instance, in the case of the middle-third Cantor set,
\beq
\tau(q) = (q-1) \frac{\ln 2}{\ln 3} + 
	i \frac{2\pi j}{\ln 3}, \,\,
	j = 0, \pm 1, \pm 2, \ldots
\label{eq:complex}
\enq
The imaginary parts of these complex roots
correspond to the period of $G$ in
(\ref{eq:bloch}) and the logarithmic period observed in the correlation 
integrals \cite{Smith}.

The reason for the surprising success of this approach, which reduces
the sometimes formidable calculation of the energy integrals
(\ref{eq:energy}) to the computation of a partition function 
(\ref{eq:partition}),
is that one has implicitly utilized the recursive structure of the
fractal distribution [encoded as (\ref{eq:bloch})].  In doing so,
one successfully captures the fact that the fractal has a 
very well defined set of real-space singularities, and equates
the determination of the spectrum of the many-body problem
to the solution of a relatively simple transcendental equation.

So far we have examined results for deterministic fractals, where
the scale invariance of (\ref{eq:bloch}) is exactly satisfied.
Consider now the case in which the fractal is not exactly self-similar,
but is only statistically self-similar, \ie, scaling functions obey
(\ref{eq:bloch}) only on average.  We shall make precise what this
averaging procedure entails (for related issues of averaging of
stochastic hierarchical processes, see \cite{Hentschel}).

For a process that generates a generalized Cantor set by replacing
each segment at level $l$ by $m$ segments at level $l+1$, we let the
length of the segments, $\ell_i (i=1, \ldots, m)$, be random variables
with random probability measures $p_i (i=1, \ldots, m)$.
For this and other random fractals, we define a new set of
correlation integrals as the expectations of (\ref{eq:correlation}),
given a probability distribution of the 
$p_i$'s and the $\ell_i$'s:
\beq
\C_q(r) = E\left[C_q(r)\right],
\enq
where $E$ denotes expectation and $C_q(r)$ is the value of the
correlation integral for a single realization of the random fractal.
A new energy integral may be defined in precisely the same fashion:
\beq
\I_q(\tau) = E\left[I_q(\tau)\right], \quad \tau < d.
\enq

It has been shown by Falconer \cite{FalconerDq} that the
probabilistic version of equation (\ref{eq:partition}) still
gives the relevant dimension spectrum, i.e, one has to solve
the expectation equation,
\beq
E\left[\sum_{i=1}^{m} \frac{p_i^q}{\ell_i^\tau}\right] = 1,
\label{eq:expectation}
\enq
to obtain the multifractal dimension spectrum.

The conditions for the existence of a unique and meaningful real
solution of (\ref{eq:expectation}) have been
studied \cite{FalconerDq,Graf,cuttreesums}, and simple extensions
of these considerations lead to the existence of well-defined
complex solutions in complete analogy with the deterministic case.

Consider then the following example of a randomized Cantor set.
At level $l$ of the recursive construction, we divide each
segment into $n$ equal segments and pick $m$ of them at random
with uniform probability.  We assign to each of the $m$ smaller
segments a measure $1/m$-th of the original segment.
To simplify the presentation, we examine only the case for
$q=2$ and do not consider the more general model of Falconer
\cite{FalconerDq}, which involves possibly non-uniform probability
distributions $p_i$ and $\ell_i$.  However, the computation may be
generalized for higher-order correlation integrals and for
non-uniform probability distributions satisfying the restrictions
outlined in \cite{FalconerDq}.

At level $l$ of this process, the energy integral is simply related
to the energy integral of the previous level:
\beq
  \I^{(l)}(\tau) = m n^{-\tau}\, \I^{(l-1)}(\tau) + \R(m,n,\tau),
\label{eq:recur}
\enq
where the superscripts denote the finite-level approximation of
the energy integral.  The function $\R$ is given by
\beq
\R(m,n,\tau) = \frac{2 n (m-1)(1-n^{\tau-2})}
{m (n-1) (\tau-1)(\tau-2)}.
\label{eq:residue}
\enq
Explicit derivation of $\R$ for this
example is given as an appendix.
Note that in (\ref{eq:recur}), the prefactor $m n^{-\tau}$ is the
expected value of the partition function for this model.
So that the energy integral of the limiting distribution is simply
\beq
\I(\tau) = \frac{\R(m,n,\tau)}{1 - m n^{-\tau}}\,.
\label{eq:2energy}
\enq

All of the singularities of $\I(\tau)$ are given by the zeros of
the denominator on the right-hand side of (\ref{eq:2energy}),
\beq
m n^{-\tau} = 1,
\label{eq:model}
\enq
as expected.  It is easily
checked that (\ref{eq:residue}) is not singular at $\tau = 1$.
%There is an additional singularity % of (\ref{eq:2energy}) 
%at $\tau = 2$, arising from the pole of (\ref{eq:residue}).
%This pole is an artifact of our model---there is a non-vanishing
%probability of adjacent segments touching each other---and 
%is not relevant.

The roots of (\ref{eq:model}) are at
\beq
\tau_j = \frac{\ln m}{\ln n} + i\frac{2\pi j}{\ln n} \,, 
\quad j = 0,\pm 1,\pm 2,\ldots
\label{eq:poles}
\enq
Thus, we expect the correlation integral to exhibit oscillations of
period $\ln n$.  By using properties of inverse Mellin transforms
\cite{Mellin}, the correlation integral can be written as
\beq
%\begin{eqnarray}
%\nonumber
%\C(r) = r^{\ln m/\ln n} \left[\frac{\alpha_0}{\tau_0}
%+ 2 \sum_{j=1}^{\infty} 
\C(r) = r^{D} \left(\frac{\alpha_0}{\tau_0} + 2 \sum_{j=1}^{\infty}
\left|\frac{\alpha_j e^{i \phi_j}}{\tau_j}\right|
\cos\left(2\pi j\frac{\ln r}{\ln n} - \phi_j\right)\right),
\label{eq:c2}
%\end{eqnarray}
\enq
where $D = \ln m/\ln n$ is the (second-order) correlation
dimension.  $\alpha_j$ and $\phi_j$ are real and are determined
by the residues at $\tau_j$ (for $j$ non-negative): The residue at
the $j$-th pole is of the form, $\alpha_j \exp(i\phi_j)$.
We propose to call the modulus of $\alpha_j e^{i\phi}/\tau_j$
(and the higher-order analogs) the lacunary amplitudes.
%Note that $\alpha_0/\tau_0$ is related to the second-order Renyi
%information, $I_2 = \ln (\alpha_0/\tau_0)$.

In Fig.\ 1, we compare the expected scaling of the correlation
integral with an ensemble average of the scaling for the case
$n = 3$ and $m = 2$.  The average is performed over several
(in this case, twelve) numerical realizations of the random
fractal (approximated at level $l = 15$).
We plot the residuals, $r^{-\ln m / \ln n} \C(r)$,
versus $r$.  The dashed line is the ensemble average, and the
solid line exhibits the expected oscillations of (\ref{eq:c2}).

\begin{figure}
%\vspace{95mm}
\psfig{file=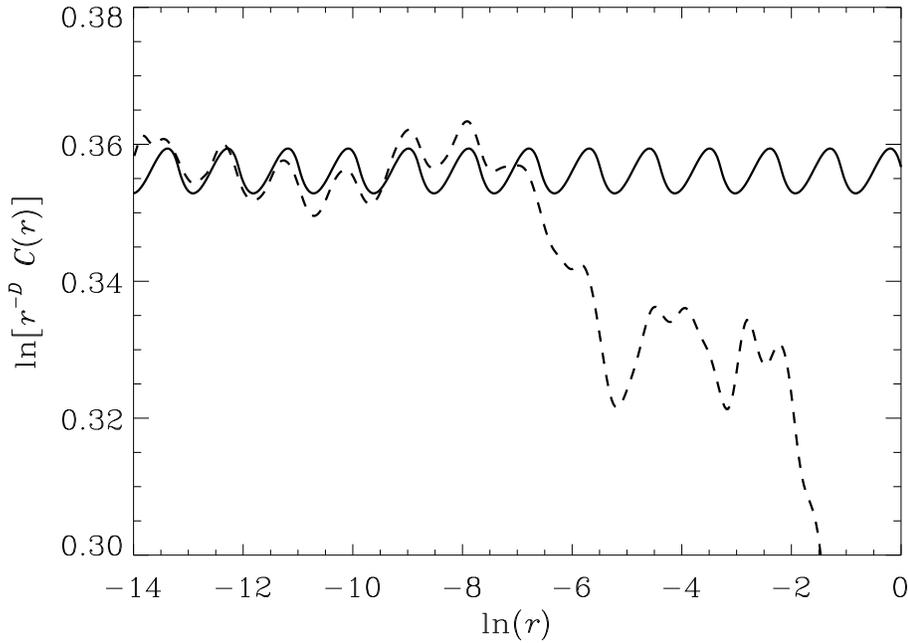,height=95mm}
\caption{Log-log plot of $r^{-D} \C(r)$ versus $r$. Dotted curve
is an ensemble average; solid curve is the expectation.}
\end{figure}

While it may seem superfluous to obtain the correlation integral
from numerical computations once it has been calculated
analytically, this exercise was interesting
since we have not performed the average over a large
ensemble.  Rather we took
the spatial average of a few
instances of a random process.
That both results were essentially the same over a number of
logarithmic periods is a rather natural self-averaging
property of many fractal objects.
The fluctuation of the ensemble average about the expectation
may be described by higher-order correlation integrals 
and will be discussed in a forthcoming publication \cite{ST}.
In the present case, the small size of the ensemble produces
disagreement with predictions at $r$ of the order of the
system size.  Furthermore, numerical resolution affects
the correlation for $r$ smaller than $e^{-13}$.

%We note that the singularity spectrum is an invariant
%of our model of random fractals (determined by the values of
%$m$ and $n$, and the probability
%distributions of the $p_i$'s and the $\ell_i$'s).
%However, the residues at each pole 
%are not universal; that is to say, the values of the residues
%may vary not only with the parameters of the model.
%%may vary not only with the parameters of the model \cite{parameters},
%%\bibitem{parameters} For instance,
%%we have specified that each of the $m$ smaller segments are picked
%%at random (with uniform probability) from the $n$ divisions of
%%the original segment.  A non-uniform probability distribution of
%%the placement of the smaller segments would lead to 
%%differently-valued residues at the very same complex poles.
%but also with different measures of correlation.
%For instance, non-universal features are exhibited in the
%oscillations of the scaling of the Fourier transform \cite{Fourier}
%and the diffraction spectrum %(or structure factor)
%\cite{Fourier,diffraction}.  While the same logarithmic
%periodicity is observed in the oscillations of the spectrum,
%the shape and phase of the oscillations are governed by
%differently-valued residues located at the roots of
%(\ref{eq:expectation}), and, hence, are non-universal.

We note that the measurement of the lacunarity spectrum is
accessible using a variety of correlations.  For instance,
we may identify these complex dimensions in the logarithmically
periodic oscillation of the Fourier transform \cite{Fourier}
and the diffraction spectrum \cite{Fourier,diffraction}.  
However, while the same logarithmic periods are observed, the 
shape and phase of the oscillations vary.  The periods
arise from the additional discrete scale invariance of the underlying
model and the numerical value of these periods are determined
by the solutions of (\ref{eq:expectation}).
The different measures
of correlation (Fourier or Mellin transforms) reveal 
differently-valued residues located at the roots of 
(\ref{eq:expectation}).

Previous studies of the inverse fractal problem, \ie, the
extraction of the (possibly stochastic)
hierarchical process from the observed
multifractal dimension spectrum, revealed ambiguities in the
standard procedure: Namely, many models can be made to fit a
given multifractal dimension spectrum \cite{Feigenbaum,Chhabra}.
The approach
described above provides the {\it maximal} characterization
of the underlying multiplicative process without additional
dynamical information.  In this way, we may further distinguish
between different hierarchical processes that give rise to fractals
with similar dimension spectra (see also \cite{Hentschel}).

We have to stress that the lacunarity spectrum does not resolve
all the inherent
ambiguities.  Many models, random or deterministic,
can be made to fit a given lacunarity spectrum.
However, using the lacunarity spectrum, we may distinguish
between processes that have the same multifractal dimension
spectrum.  What the lacunarity spectrum reveals is the
possible additional
discrete scale invariance which are not furnished by the
previous attempts to characterize fractal systems.
%However,
%the lacunarity spectrum contains information about the possible
%discrete scale invariance of these systems, in addition to the
%information given by the usual multifractal dimension spectrum.
Furthermore, non-universal information is recovered from studying
correlations (\ref{eq:correlation}) and energies (\ref{eq:energy}).

How does this work bear on fractal sets generated by low-order
(deterministic or stochastic) dynamical systems?  Preliminary
work \cite{ST} suggests that the complex solutions of the
Lyapunov partition function (a dynamical analog of
(\ref{eq:partition}) and (\ref{eq:expectation}), see \cite{Lyap})
describes the anomalous scaling observed in simulations (see,
for instance, \cite{Badii}).  In addition, the fact
that the lacunarity spectrum can be calculated from a 
partition function (\ref{eq:partition}) immediately implies
that periodic orbit expansions \cite{periodic} can be used to
calculate the spectrum.

As for fractals generated by systems governed by large numbers
of degrees of freedom (as featured most prominently in phenomena
modeled by diffusion limited aggregation and in inhomogeneities
of highly turbulent flows), this generalized
multifractal description complements the traditional
views.  Studies thus far have taken the position that the
deviations from strict power-law scaling are anomalous and, hence,
have focused on establishing possible causes for this apparent 
deviation (an example being inertial range intermittency
\cite{PV,intermittent} in turbulent fluids).  In our approach,
log-periodic deviations may be accommodated rather naturally
(see also \cite{Smith,Novikov}).  Future
efforts will be directed towards the description of turbulent
intermittency using the analysis followed in this paper.

%\n{\bf Acknowledgment}

We gratefully acknowledge J. Fournier, R. Rosner, 
A. Sornborger, and E. Spiegel for helpful conversations.
We thank R. Ball for bringing his work with R. Blumenfeld
to our attention.  We also wish to thank the anonymous
referee for useful suggestions.
This work was completed while F.~J.~S. was a Rosenbaum fellow at
the Newton Institute.  L.~T. is supported by the U.~K. Particle
Physics and Astronomy Research Council.

%\b\b\b
%\newpage

%\begin{appendix}

\appendix
\n{\bf Appendix}

\renewcommand{\theequation}{A\arabic{equation}}
\setcounter{equation}{0}
\renewcommand{\thefigure}{\arabic{figure}}
\setcounter{figure}{1}

In this Appendix we present the details leading to the explicit
formulas of the energy integral as given by (\ref{eq:residue})
and (\ref{eq:2energy}).

We need to consider first the behavior of the measure
upon averaging. Since we have assigned an equal probability to
every possible case of segmentation, the expectation value
for the density, $\rho(x)dx=d\mu(x)$, is uniform, i.e.,
\begin{equation}
        E\left[\rho(x)\right]=1.
\end{equation}

To be able to perform the required multiple integrals, we need to
evaluate the expectation of products of densities.
Sufficient information about the joint distribution of these densities
is obtained by considering one step in the recursive construction.
After one such step, the interval $L^{0}=[0,1]$ is divided into $n$
subintervals $L^{1}_{i}, i=1, 2,\ldots, n$
from which $m$ subintervals will be chosen randomly.
The set of all points $(x,y)$ that belong to
the subintervals $L^{1}_{i}$ and $L^{1}_{j}$, respectively,
form a square of size $n^{-1}\times n^{-1}$
in the $x$-$y$ plane. Such a square will be labelled $(i,j)$, as shown in
Figure 2. The shaded region $R$ corresponds to those cases in which
$x$ and $y$ are in disjoint intervals, i.e., $i \ne j$.

\begin{figure}
%\vspace{80mm}
{\hskip 30 mm{\psfig{file=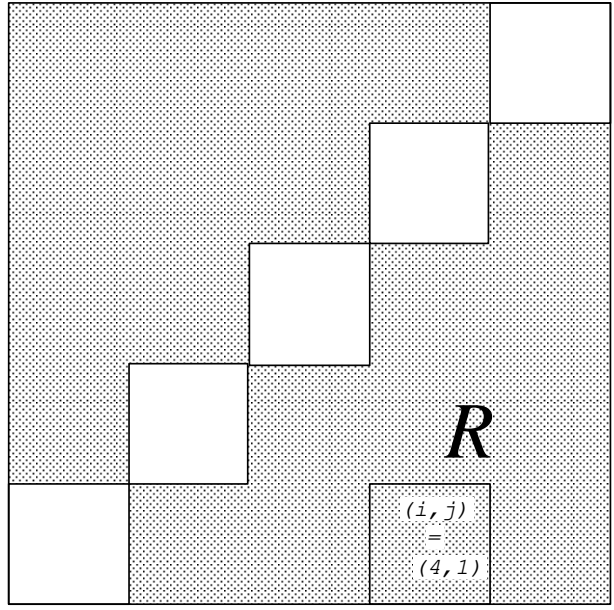,height=80mm}}}
\caption{Different regions of integration % recurrence relation
for the energy integral (\ref{eq:recur}).
Shaded regions (labelled by $R$) contribute to the function
$\R(m,n,\tau)$; white regions contribute to $\I^{(l)}(\tau)$.
Labels of individual squares, $(i,j)$, are explained in the text.
Shown here is a case for $n = 5$.}
\end{figure}

Consider now one of the shaded squares, say $(i,j)$. The probability
that the intervals $L^{1}_{i}$ and $L^{1}_{j}$ are indeed chosen is simply
 ${n-2\choose m-2}/{n\choose m}$.
In this case, each of the intervals will support a measure of
total mass $1/m$.  Furthermore, since
segmentation for each of these intervals proceeds
uncorrelated, we have
\bearr
\nonumber
E\left[\rho(x)\rho(y)\right] & = & \frac{{n-2\choose m-2}}{{n\choose m}}
E_{i}\left[\rho(x)\right] E_{j}\left[\rho(y)\right]\\
& = & \frac{m(m-1)}{n(n-1)}\frac{n^{2}}{m^{2}}, \quad (x,y)\in (i,j),
\enarr
where $E_{k}$ denote expectation conditioned to the event that
segment $k$ is indeed chosen.

Next, we consider the diagonal squares. These squares will contribute
to the energy integral with probability ${n-1\choose m-1}/
{n\choose m}=m/n$. The process of segmentation for a subinterval is
identical to that of the original interval. Therefore, if $L^{1}_{i}$
is chosen, the average of the product of densities
$\rho(x)\rho(y)$ restricted to this interval is identical
to that of the original interval up to rescaling.
The rescaling matches $x$ from the $L^{1}_{i}$ segment with
the point $x'=nx-i-1$ in $L^{0}$.
We have, for $x,y \in L^{1}_{i}$,
\bearr
\nonumber
E\left[ \rho(x)\rho(y) \right]
& = & (m/n) E_{i}\left[ \rho(x)\rho(y) \right]\\
& = & (m/n)(1/m)^{2} E\left[ \rho(x')\rho(y') \right]
\enarr
Note also that $|x-y|=(1/n)|x'-y'|$.
Thus the contribution to the energy integral from each of the diagonal
squares is proportional to the overall expectation of the energy integral
\begin{equation}
E\left[\int_{(i,i)}\frac{d\mu(x)d\mu(y)}{|x-y|^{\tau}}\right]=
(m/n)(1/m)^{2} n^{\tau}\, {\I}_2(\tau)
\end{equation}

Summing over all squares we obtain relation (\ref{eq:recur}) where
${\R}(m,n,\tau)$ can now be identified with the expectation value
of the energy integral restricted to the shaded region.

To simplify the calculation of ${\R}(m,n,\tau)$ we note that
\begin{equation}
E\left[ \int_{R}\frac{d\mu(x)d\mu(y)}{|x-y|^{\tau}}\right] =
\frac{m(m-1)}{n(n-1)}\left(\frac{n}{m}\right)^{2}
\int_{R}\, \frac{dx dy}{|x-y|^{\tau}}
\end{equation}
Furthermore, the last integral satifies
\begin{equation}
\int_{0}^{1}\int_{0}^{1}\, \frac{dx dy}{|x-y|^{\tau}}=
\int_{R}\, \frac{dx dy}{|x-y|^{\tau}} +
n^{s-1}\int_{0}^{1}\int_{0}^{1}\, \frac{dx dy}{|x-y|^{\tau}}
\end{equation}
which readily gives the final result
\begin{equation}
R(m,n,\tau) = \frac{2 n (m-1)(1-n^{\tau-2})}
{m (n-1) (\tau-1)(\tau-2)}.
\end{equation}

%\end{appendix}

%\newpage

\end{document}